\documentstyle[preprint,floats,prd,aps,eqsecnum,epsf,12pt]{revtex}

\begin{document}

\preprint{\vbox{\hbox{July 1998} \hbox{IFP-762-UNC} 
\hbox{KEK Preprint 98-106} }}
\title{%
Chiral Phase Transition of QCD at Finite Temperature and Density
from Schwinger-Dyson Equation
}
\author{Masayasu Harada\footnote{%
Electronic address: {\tt harada@physics.unc.edu}}}
\address{Department of Physics and Astronomy, 
University of North Carolina,\\
Chapel Hill, NC 27599-3255.}
\author{Akihiro Shibata\footnote{%
Electronic address: {\tt ashibata@post.kek.jp}}}
\address{Computing Research Center, High Energy Accelerator Research
Organization (KEK),\\ Tsukuba 305, Japan}
\maketitle

\begin{abstract}
We study the chiral phase transition of QCD at finite temperature and
density by numerically solving Schwinger-Dyson equation for the quark
propagator with the improved ladder approximation in the Landau gauge.
Using the solution we calculate a pion decay constant from a
generalized version of Pagels-Stokar formula.  Chiral phase transition
point is determined by analyzing an effective potential for the quark
propagator.  We find solutions for which chiral symmetry is broken
while the value of the effective potential is larger than that for the
chiral symmetric vacuum.  These solutions correspond to meta-stable
states, and the chiral symmetric vacuum is energetically favored.  We
present a phase diagram on the general temperature--chemical potential
plane, and show that phase transitions are of first order in wide
range.
\end{abstract}

\section{Introduction}

In QCD an approximate chiral symmetry exists at the Lagrangian level
and it is spontaneously broken by the strong gauge interaction.
Accordingly approximate Nambu-Goldstone (NG) bosons appear, and pions
are regarded as the NG bosons.  In hot and/or dense matter, however,
quark condensates melt at some critical point, and the chiral symmetry
is restored (for recent reviews, see, e.g., Ref.~\cite{finiteT}).
To study the chiral phase transition 
we need non-perturbative treatment.

Schwinger-Dyson Equation (SDE) is a
powerful tool for studying the qualitative structure of chiral
symmetry breaking of QCD.  By suitable choices of the running 
coupling, in which the asymptotic freedom is incorporated, the SDE
provides a non-trivial solution for the mass function.  The behavior
of the mass function at high energy is consistent with that given
by the operator product expansion technique (for a review, see,
e.g., Ref.~\cite{Kugo:BSrev}).  The SDE is understood as a stationary
condition of an effective potential for the quark propagator, and the
chiral broken solution is energetically favored by the effective 
potential\cite{DM64,CJT,Haymaker,BHK:BS}.

The SDE was applied for studying the chiral phase transition of QCD at
finite temperature and/or density~\cite{Ka85,CGS86,%
GGKPS86,Akiba87,Ka88,CKV88,BCDGP:PRD41,BCDGP:PLB240,OS91,%
KY95,Taniguchi-Yoshida,LLN}.
In many of those the chiral phase transition point was determined from
the point where the mass function (or pion decay constant) vanished
without studying an effective potential.
In Ref.~\cite{BCDGP:PRD41} an effective potential for the mass
function was analyzed by using an approximate form of momentum
dependence of the mass function.
It was shown that there were meta-stable states where the chiral
symmetry was broken while the energy was bigger than that for the
symmetric vacuum.
It is interesting to study whether this kind of meta-stable
state is derived by fully solving the SDE.

In this paper we study the chiral phase transition in QCD 
at finite temperature and/or density  
by numerically solving  the SDE 
using Matsubara formalism.
As an order parameter we
calculate a pion decay constant using a generalized version of
Pagels-Stokar formula~\cite{Pagels-Stokar}.  
We calculate  the value of the effective potential at the solution 
of the SDE to determine the chiral phase transition points.
We show that meta-stable states actually exist in wide range.
The order of phase transition are determined by comparing the order
parameter with the effective potential. 
The phase transition is of the first order for wide range at 
$T\neq0$ and $\mu\neq0$.

This paper is organized as follows.  An effective potential for the
quark propagator is introduced at finite temperature and/or density in
section~\ref{sec:2}.  The SDE is derived as a stationary condition of
the potential.
In section~\ref{sec:fpi} a formula for calculating a pion decay
constant is presented.
Section~\ref{sec:numerical} is a main part of this paper, where
numerical solutions of the SDE are shown.
Values of the pion decay constant and the effective potential are
calculated from the solutions, and the phase structure is studied.
Finally a summary and discussion are
given in section~\ref{sec:discussion}.

\section{Effective Potential and Schwinger-Dyson Equation}
\label{sec:2}

At zero temperature
the effective action for
the quark propagator $S_F$ is given by~\cite{CJT,BHK:BS}
\begin{equation}
\Gamma[S_F] = i \mbox{Tr}\,\mbox{Ln}\,S_F
- \mbox{Tr} \left[ i \partial\kern-5.5pt\mbox{\it/} S_F \right]
- i \kappa_{\rm 2PI}[ S_F ]
\ ,
\end{equation}
where $\kappa_{\rm 2PI}$ stands for the two particle irreducible (with
respect to quark-line) diagram contribution.
In this paper we take the first order approximation for 
$\kappa_{\rm 2PI}$, 
in which one gluon exchange graph contribute, 
\begin{equation}
\kappa_{\rm 2PI}[S_F] =
- \frac{N_{\rm f} N_{\rm C} C_2}{2} \int d^4 x d^4 y g^2 
\mbox{tr}\, 
\left[ S_F(x-y) i \gamma_\mu S_F(y-x) i \gamma_\nu \right]
D^{\mu\nu}(x-y)
\ ,
\end{equation}
where the number of flavor ($N_{\rm f}$) and of
 color ($N_{\rm C}$) and
the second Casimir of SU($N_{\rm C}$) ($C_2$) appear since
we factor color and flavor indices from the quark and gluon
propagators.
In this expression, the quark propagator $S_F$ carries bispinor
indices and the gluon propagator $D^{\mu\nu}$ carries Lorenz
indices.
{}From the above effective action the effective potential for $S_F$ is
obtained by the usual procedure.
In momentum space it is expressed as
\begin{eqnarray}
V[S_F] &=&
\int \frac{d^4p}{i(2\pi)^4}
\left\{ 
  \mbox{\rm ln}\,\mbox{\rm det}\,\left[ S_F(p) \right]
  + \mbox{\rm tr}\, \left[ p\kern-5.5pt\mbox{\it/} i S_F(p) \right]
\right\}
\nonumber\\
&&
+ \frac{1}{2} 
\int \frac{d^4p}{i(2\pi)^4} \int \frac{d^4k}{i(2\pi)^4} C_2 g^2(p,k)
\mbox{\rm tr}\, \left[ i S_F(p) \gamma_\mu i S_F(k) \gamma_\nu \right]
i D^{\mu\nu}(k-p)
\ ,
\label{eff pot}
\end{eqnarray}
where an overall factor $N_{\rm C}N_{\rm f}$ is dropped.
The running coupling $g^2(p,k)$ is introduced to improve 
Schwinger--Dyson Equation (SDE) to include aymptotic freedom of QCD.
An explicit form of the running coupling will be given below.  In this
paper we take the Landau gauge for the gluon propagator: 
\begin{equation}
i D^{\mu\nu}(l) = \frac{1}{l^2} 
\left[ 
  g^{\mu\nu} - \frac{l^\mu l^\nu}{l^2}
\right] \ .
\label{gluon prop}
\end{equation}

General form of the full quark propagator is expressed as
\begin{equation}
i S_F^{-1}(p) = A(p) p\kern-5.5pt\mbox{\it/} - B(p) \ .
\end{equation}
SDE for the quark propagator $S_F$ is given as a stationary condition
of the above effective potential\cite{CJT} (See also,
Ref.~\cite{BHK:BS}). 
The SDE gives coupled equations for $A$ and $B$.
If the running coupling $g(p,k)$ is a function in $p^2$ and $k^2$,
{\it i.e.}, it does not depend on $p\cdot k$, it is easy to show that 
$A(p^2)=1$ is a solution of the SDE in the Landau gauge with the
ladder approximation.

Let us go to non-zero temperature and density.  In the imaginary time
formalism\cite{Matsubara} the partition function is calculated by the
action given by (See, for reviews, e.g.,
Refs.~\cite{Toimele,Kapusta}.)
\begin{equation}
S = \int^{1/T}_0 d \tau \int d^3x
\left[ {\cal L}_{\rm QCD} + \mu \overline{\psi}\gamma^0\psi \right]
\ ,
\end{equation}
where $T$ and $\mu$ denote temperature and chemical potential
associated with quark (baryon) number density, and 
${\cal L}_{\rm QCD}$ takes the same form as QCD Lagrangian at
$T=\mu=0$.  
{}From the above action we obtain an effective potential for
the quark propagator $S_F$ similar to the one in 
Eq.~(\ref{eff pot}):
\begin{eqnarray}
\lefteqn{
V[S_F] =
T \sum_u
\int\!\frac{d^3\vec{p}}{(2\pi)^3}
\left\{ 
  \mbox{\rm ln}\,\mbox{\rm det}\,\left[ S_F(p) \right]
  + \mbox{\rm tr}\, \left[ p\kern-5.5pt\mbox{\it/} i S_F(p) \right]
\right\}
}
\nonumber\\
&&
+ \frac{1}{2} T^2 \sum_{u,v}
\int\!\frac{d^3\vec{p}}{(2\pi)^3}
\int\!\frac{d^3\vec{k}}{(2\pi)^3} C_2 g^2(p,k)
\,\mbox{\rm tr} \left[ i S_F(p) \gamma_\mu i S_F(k) \gamma_\nu \right]
i D^{\mu\nu}(k-p)
\ ,
\label{eff pot T}
\end{eqnarray}
where
\begin{eqnarray}
&& p_0 = i u + \mu \ , \quad u = (2 m + 1) \pi T \ , \nonumber \\
&& k_0 = i v + \mu \ , \quad v = (2 n + 1) \pi T \ , 
\quad (n, m = \mbox{\rm integer})\ ,
\end{eqnarray}
and $\sum_u$ and $\sum_v$ imply summations over $m$ and $n$,
respectively.

Argument of the running coupling should be taken as $(k-p)^2$
for preserving the chiral symmetry~\cite{Kugo-Mitchard}.
However, as is shown in Ref.~\cite{Kugo-Mitchard}, angular average,
{\it i.e.}, the running coupling is a function in 
$-p_0^2+\vert\vec{p}\vert^2-k_0^2+\vert\vec{k}\vert^2$, is a good
approximation at $T=\mu=0$.
If we naively extend this approximation, the running coupling will
depend on $\mu$. 
However, the running coupling should not depend on $\mu$.
Then in this paper we use the approximation where the running coupling
is a function in 
$-(p_0-\mu)^2+\vert\vec{p}\vert^2-(k_0-\mu)^2+\vert\vec{k}\vert^2$.
The explicit form of the running coupling 
is~\cite{Aoki-Bando-Kugo-Mitchard-Nakatani}
\begin{eqnarray}
g^2(p,k) &=& \left(4\pi\right)^2
\frac{3}{11N_{\rm C} - 2N_{\rm f}}\,
\overline{\lambda} 
\left( \mbox{ln}\left( \frac{u^2+x^2+v^2+y^2}{\Lambda_{\rm qcd}^2}
\right)\right) \ ,
\nonumber\\
\overline{\lambda}(t) &=&
\left\{ \begin{array}{ll}
\displaystyle
\frac{1}{t}  & \mbox{if} \ t_{\rm F} < t \\
\displaystyle
\frac{1}{t_{\rm F}} +
\frac{
  \left(t_{\rm F}-t_{\rm C}\right)^2 - \left(t-t_{\rm C}\right)^2
}{
  2 t_{\rm F}^2 \left(t_{\rm F}-t_{\rm C}\right)
}
& \mbox{if} \ t_{\rm C} < t < t_{\rm F} \\
\displaystyle
\frac{1}{t_{\rm F}} +
\frac{\left(t_{\rm F}-t_{\rm C}\right)}{2 t_{\rm F}^2}
& \mbox{if} \ t < t_{\rm C} \\
\end{array}
\right.
\ ,
\label{running coupling}
\end{eqnarray}
with
\begin{equation}
x \equiv \vert\vec{p}\vert \ , \ \ y \equiv \vert\vec{k}\vert \ .
\label{xy}
\end{equation}
$\Lambda_{\rm qcd}$ is a scale where the one-loop running coupling
diverges, and the value of $\Lambda_{\rm qcd}$ will be determined
later from the infrared structure of the present analysis.\footnote{%
The usual $\Lambda_{\rm QCD}$ is determined from the ultraviolet
structure.
}
$t_{\rm F}$ and $t_{\rm C}$ are parameters introduced to
regularize the infrared behavior of the running coupling.
In the numerical analysis below, since the dominant part of the mass
function lies below the threshold of charm quark, we will take 
$N_{\rm f}=3$ and $N_{\rm C}=3$.
Moreover, we use fixed values $t_{\rm F}=0.5$ and 
$t_{\rm C}=2.0$~\cite{Kugo:BSrev}.

General form of the full quark propagator,
which is invariant under the Parity-transformation, 
is given by\cite{Ka85,Ka88,CKV88,OS91}
\begin{equation}
i S_{\rm F}^{-1}(p) = A(p) p\kern-5.5pt\mbox{\it/}
- B(p) + C(p) p_0 \gamma^0 
- \frac{i}{2} \left[ \gamma^0 \,,\, p\kern-5.5pt\mbox{\it/}
 \right] \, D(p)
\ .
\label{prop inv}
\end{equation}
Here $A$ $\sim$ $D$ are functions in $p_0$ and
$\vert\vec{p}\vert$ (as well as $T$ and $\mu$), although we write
$A(p)$, etc.

It should be noticed that 
the effective potential in 
Eq.~(\ref{eff pot T}) is 
invariant under the tranfromation:
\begin{equation}
S_F(p_0,\vec{p}) \longrightarrow S_F^\prime(p_0,\vec{p}) = 
\left( i \gamma^1\gamma^3\gamma^0 \right)
S_F^T(p_0,-\vec{p})
\left( i \gamma^0\gamma^1\gamma^3 \right) \ ,
\end{equation}
where $S_F^T$ implies transposition of $S_F$ in spinor 
space.\footnote{
In our convention, $(\gamma^0)^{\ast} = \gamma^0$,
$(\gamma^1)^{\ast} = \gamma^1$ and $(\gamma^3)^{\ast} = \gamma^3$,
while $(\gamma^2)^{\ast} = -\gamma^2$.}
At $T=\mu=0$ this is nothing but the time reversal symmetry.
It is natural to expect that this ``time reversal'' symmetry is not
spontaneously broken.
Existence of this ``time reversal'' leads $D(p)=0$.

Moreover, at $\mu=0$ the effective potential is invariant under the
following ``charge conjugation'':
\begin{equation}
S_F(p) \longrightarrow
S_F^\prime (p) = - (i\gamma^2\gamma^0) S_F^T(-p) (i\gamma^2\gamma^0)
\ .
\end{equation}
We also assume this symmetry is not spontaneously broken, and it
implies that all the scalar functions $A$, $B$ and $C$ (as well as
$D$) are even functions in $p_0=i u$.
As we mentioned above, at $T=\mu=0$, of course $C(p)=0$, and $A(p)=1$
in the Landau gauge. 
$C(p)\neq0$ as well as $A(p)\neq1$ does not imply chiral symmetry
breaking, and non-zero $B(p)$ is the only signal of the breaking in
the present analysis.
Then we consider $A(p)-1=C(p)=0$ as an approximate
solution at general $T$ and $\mu$.
We note that non-zero $D(p)$ also breaks chiral symmetry, however 
this 
term do not contribute to local order parameter of ciral symmetry,
$\langle\overline{\psi}\psi\rangle = -\int \mbox{tr} S_F$. 
We choose vacuum of ``time reversal'' invariance as discussed 
above.\footnote{Acctually, in the numerical calcuration of SDE 
in bifercation apploximation, no $D(p)\not=0$ solution is obtained}

As in the case at $T=\mu=0$, the SDE is given as a stationary
condition of the effective potential (\ref{eff pot T}):
\begin{equation}
i S_{\rm F}^{-1}(p) - p\kern-5.5pt\mbox{\it/} = T \sum_v \int 
\frac{d^3 \vec{k}}{\left(2\pi\right)^3} C_2 g^2(p,k) \gamma_\mu
i S_{\rm F}(k_f) \gamma_\nu i D^{\mu\nu}(k-p) \ .
\label{SD: finite T}
\end{equation}
By taking a trace and performing the three-dimensional angle
integration (Note that the present form of the running coupling does
not depend on the angle.), we obtain a self-consistent equation for
$B$:
\begin{equation}
B(p) = K(p,k) \ast \frac{B(k)}{B^2(k)-k^2}
\equiv 
T \sum_v \int \frac{dy y^2}{2\pi^2}
K(p,k) \frac{B(k)}{B^2(k)-k^2}
\ ,
\label{SD: B}
\end{equation}
where 
\begin{equation}
K(p,k) \equiv \frac{3}{2} C_2 g^2(p,k) \times
\frac{1}{2x y} \ln \left(
\frac{(u-v)^2 + (x+y)^2}{(u-v)^2 + (x-y)^2}
\right)
\ .
\end{equation}

In general cases $B$ is a complex function.
By using the fact that $K(p,k)$ is a real function,
it is easily shown that $B^{\ast}(p)$ satisfies the same equation
as $B(p^{\ast})$ does, where 
$p^{\ast\mu} = (p^{0\ast},\vec{p}) = (- i u + \mu, \vec{p})$. 
So these two are equal to each other up to sign,
$B^{\ast}(p) = B(p^{\ast})$ or $B^{\ast}(p) = - B(p^{\ast})$.
Since $B$ is an even function in $p_0$ at $\mu=0$, the choice of
positive (negative) sign implies that $B$ is real (pure imaginary) at
$\mu=0$.
$B$ should be real at $\mu=0$, then
\begin{equation}
B^{\ast}(p) = B (p^{\ast}) \ .
\label{rel: B}
\end{equation}

After a solution of the SDE (\ref{SD: finite T}) is substituted with
the approximate solution $A(p)-1=C(p)=0$ into Eq.~(\ref{eff pot T}),
the effective potential becomes
\begin{eqnarray}
\overline{V}[B_{\rm sol}] 
&\equiv&
V[B_{\rm sol}] - V[B =0] 
\nonumber\\
&=& \frac{2}{\pi^2} T \sum_u \int dx x^2
\left[ 
  -\ln \left( \frac{B^2_{\rm sol}(p) - p^2}{-p^2} \right)
  + \frac{B^2_{\rm sol}(p)}{B^2_{\rm sol}(p)-p^2}
\right]
\ ,
\label{eff pot sol}
\end{eqnarray}
where $B_{\rm sol}$ denotes a solution of Eq.~(\ref{SD: B}).
This value of the effective potential is understood as the energy
density of the solution.
So the true vacuum should be determined by studying the value of the
effective potential.  When the value of $\overline{V}[B_{\rm sol}]$
for a nontrivial solution $B_{\rm sol}$ is negative, the chiral broken
vacuum is energetically favored.  Positive 
$\overline{V}[B_{\rm sol}]$, however, implies that the chiral
symmetric vacuum is the true vacuum.

\section{Pion Decay Constant}
\label{sec:fpi}

At zero temperature and zero chemical potential
the pion decay constant is defined by the matrix
element of the axial-vector current between the vacuum and one-pion
state:
\begin{equation}
\langle 0 \vert J_{5\mu}^a(0) \vert \pi^b(q) \rangle
= i \delta^{ab} q_\mu f_\pi
\ ,
\label{def: fpi0}
\end{equation}
where $a$ and $b$ denote iso-spin indices and
$J_{5\mu}^a = \overline{\psi} \gamma_\mu \gamma_5 T^a \psi$
is an axial-vector current with 
$T^a$ being a generator of SU($N_{\rm f}$).
At finite temperature and density there are two distinct pion decay 
constant\cite{Pisarski-Tytgat} according to space-time symmetry of
matirix element,which are defined by
\begin{equation}
\langle 0 \vert J_{5\mu}^a(0) \vert \pi^b(q) \rangle_{T,\mu}
= i \delta^{ab}
\left[ 
  V_\mu (V\cdot q) f_{\pi L} + (g_{\mu\nu}-V_\mu V_\nu) q^\nu 
  f_{\pi T}
\right]
\ ,
\label{def: fpi}
\end{equation}
where $V^\mu = (1,\vec{0})$ is a vector of medium.
In this expression $f_{\pi L}$ and $f_{\pi T}$ are defined in the
zero momentum limit, $q\rightarrow 0$~\cite{Pisarski-Tytgat}.

The above pion decay constants are expressed 
by using the amputated pion BS amplitude $\hat{\chi}$ as
\begin{eqnarray}
&&
V_\mu (V\cdot q) f_{\pi L} + (g_{\mu\nu}-V_\mu V_\nu) q^\nu f_{\pi T}
\nonumber\\
&& \qquad
= - \frac{N_{\rm C}}{2} T \sum_{u}
\int \frac{d^3p}{(2\pi)^3} \mbox{tr} \,
\left[
  \gamma_\mu \gamma_5 i S_F(p+q/2) \hat{\chi}(p;q) i S_F(p-q/2)
\right]
\ .
\label{eq: fpi BS}
\end{eqnarray}
By using the current conservation, it is shown that the pion momentum
$\vec{q}$ and the pion energy $q_0$ satisfy the dispersion 
relation~\cite{Pisarski-Tytgat}
\begin{equation}
q_0^2 = \frac{f_{\pi T}}{f_{\pi L}} 
\left\vert \vec{q} \right\vert^2 \ .
\label{on shell}
\end{equation}

It is straightforward to obtain the pion decay constant
if we have explicit expresstion of the  amputated pion BS 
amplitude.  Although it is generally quite difficult to obtain
the BS amplitude, one can determine it in $q\rightarrow0$ limit,
$\hat{\chi}(p;q=0)$,
from the chiral Ward-Takahashi identity
\begin{equation}
i q^\mu \Gamma_{5\mu}^a \left( p-q/2 , p+q/2 \right)
= S_F^{-1}\left( p-q/2 \right) T_a \gamma_5
+ T_a \gamma_5 S_F^{-1}\left( p+q/2 \right)
\ ,
\label{axial WT}
\end{equation}
where $\Gamma_{5\mu}^a$ is the axial vector--quark--antiquark vertex
function, and we have suppressed color indices.
(Note that $q^0$ in Eq.~(\ref{axial WT}) is independent of $\vec{q}$, 
{\it i.e.}, they do not generally satisfy the pion on-shell dispersion
relation (\ref{on shell}).)
In the zero momentum limit, $\vec{q}$, $q^0\rightarrow0$
(the on-shell limit of the pion),
$\Gamma_{5\mu}^a$ is dominated by the pion-exchange contribution:
\begin{equation}
i \Gamma_{5\mu}^a \left( p-q/2 , p+q/2 \right) 
\mathop{\longrightarrow}_{\vec{q},q^0\rightarrow0}
i \left[ 
  V_\mu (V\cdot q) f_{\pi L} + (g_{\mu\nu}-V_\mu V_\nu) q^\nu 
  f_{\pi T}
\right]
\frac{1}{q_0^2 - \omega_q^2} \hat{\chi}(p,0) T_a
\ ,
\label{pion dominance}
\end{equation}
where $\omega_q^2 = \left(f_{\pi T}/f_{\pi L}\right)
\vert\vec{q}\vert^2$.
Substituting
Eq.~(\ref{pion dominance}) into  Eq.~(\ref{axial WT})
and taking the zero momentum 
limit, we find
\begin{equation}
\hat{\chi}(p;0) = \frac{2}{f_{\pi L}} B(p) \gamma_5 \ ,
\label{chi B}
\end{equation}
where we have used $D(p)=0$ in $S_F(p)$.

The approximation adopted by Pagels and 
Stokar~\cite{Pagels-Stokar} is essentially neglecting derivatives
of $\hat{\chi}$ in the zero momentum limit
(See, for reviews, e.g., Refs.~\cite{Kugo:BSrev,Tanabashi:DSB91}):
\begin{equation}
\lim_{\vec{q}\rightarrow0}
\frac{\partial}{\partial\vert\vec{q}\vert}
\hat{\chi}(p;q)
= 
\lim_{\vec{q}\rightarrow0}
\frac{\partial}{\partial q_0}
\hat{\chi}(p;q)
= 0 \ .
\end{equation}
This approximation is same as replacing $\hat{\chi}(p;q)$ in 
Eq.~(\ref{eq: fpi BS}) with $\hat{\chi}(p;0)$, and reproduces the
exact value of $f_\pi$ at $T=0$ in the ladder approximation within a
small error~\cite{Kugo:BSrev}.  Then we
use the same approximation at finite $T$ and $\mu$ in this paper.
Replacing $\hat{\chi}(p;q)$ with $\hat{\chi}(p;0)$ in
Eq.~(\ref{eq: fpi BS}), differentiating Eq.~(\ref{eq: fpi BS}) with
respect to $q^\alpha$ and taking $q\rightarrow0$ limit,
we obtain
\begin{eqnarray}
&&
i\delta^{ab}[V_\mu V_\alpha f_{\pi L} +
(g_{\mu\alpha} -V_\alpha V_\mu) f_\pi ]
\nonumber\\
&&\hspace{0.5cm}=
- \frac{N_{\rm C}}{2} 
T \sum_{u}
\int \frac{d^3p}{(2\pi)^3} \mbox{tr} \,
\Biggl[
  \gamma_\mu \gamma_5 
  \left\{
    \frac{i}{2} \frac{S_F(p)}{\partial p^{\alpha}}
    \hat{\chi}(p;0) i S_F(p)
    + i S_F(p) \hat{\chi}(p;0) 
    \frac{i}{2} \frac{S_F(p)}{\partial p^{\alpha}}
  \right\}
\Biggr]
\ .
\label{fpi L BS0}
\end{eqnarray}
Here we have incorporated the on-shell dispersion relation
(\ref{on shell}).
Then substituting Eq.~(\ref{chi B}) into Eq.~(\ref{fpi L BS0})
and taking $V_\mu V_\alpha$ part, we find
\begin{equation}
f_{\pi L}^2 = 4 N_{\rm C} T \sum_{u} \int \frac{dx x^2}{2\pi^2} 
\frac{
  B(p)\left( B(p) - p_0 \frac{\partial B(p)}{\partial p_0}\right)
}{
  \left( B^2(p) - p^2 \right)^2
} \ ,
\label{fpi L}
\end{equation}
where we performed the three-dimensional angle integration.
If the mass function $B(p)$ is a function in $p^2=p_0^2 - \vec{p}^2$,
this agrees with the formula derived in Ref.~\cite{BCDGP:PLB240}.
We also obtain a similar 
formula for $f_{\pi T}$ by selecting the 
$(g_{\mu\alpha}-V_\mu V_\alpha) $ part:
\begin{equation}
f_{\pi T} f_{\pi L} = 4 N_{\rm C}
T \sum_{u} \int \frac{dx x^2}{2\pi^2} 
\frac{
  B(p)\left( B(p) - \frac{x}{3}\frac{\partial B(p)}{\partial x}\right)
}{
  \left( B^2(p) - p^2 \right)^2
} \ ,
\label{fpi T}
\end{equation}
where $f_{\pi L}$ in the left hand side appears from the
normalization of $\hat{\chi}$ (see Eq.~(\ref{chi B})).

It is convenient to define a space-time 
averaged pion decay constant by
\begin{eqnarray}
   f_\pi^2 &\equiv& f_{\pi L}\frac{1}{4}
g^{\mu \alpha}\frac{\partial}{\partial q^\alpha}
\left( V_\mu (V\cdot q) f_{\pi L} 
+ (g_{\mu\nu}-V_\mu V_\nu) q^\nu f_{\pi T}\right) \nonumber\\
&=& \frac{1}{4}\left( f_{\pi L}^2 + 3f_{\pi T}f_{\pi L} \right) \ .
\label{def: fpi 2}
\end{eqnarray}
This expression aggrees with Pagels-Stokar
formula~\cite{Pagels-Stokar} at
$T=0$ and $\mu=0$ after replacing the integral variables.

In the formula (\ref{fpi L})
we need a derivative of $B(p)$ with
respect to $p_0$.  However what is obtained by solving the SDE
(\ref{SD: B}) is a function in discrete $u=(2n+1)\pi T$.  To obtain
the derivative an ``analytic continuation'' in $u$ from discrete
variable to continuous variable is needed.  This ``analytic
continuation'' is done by using the SDE (\ref{SD: B}) itself:
\begin{equation}
p_0 \frac{\partial B(p)}{\partial p_0} = 
p_0 \frac{\partial K(p,k)}{\partial p_0} \ast 
\frac{B(k)}{B^2(k)-k^2} \ .
\end{equation}

\section{Numerical Analysis}
\label{sec:numerical}

In this section we will numerically solve the SDE (\ref{SD: B})
and calculate the effective potential and the pion decay constant
$f_{\pi}(T)$ 
defined in Eq.~(\ref{def: fpi 2})
for the cases 1) $T=0$ and $\mu\neq0$, 
2) $T\neq0$ and $\mu=0$ and 3) $T\neq0$ and $\mu\neq0$, separately.
The essential parameters in the present analysis are 
$\Lambda_{\rm qcd}$ and $t_F$ in the running coupling 
(\ref{running coupling}).  In this paper we fix $t_{\rm F}=0.5$ and
the value of $\Lambda_{\rm qcd}$ is determined by calculating the pion
decay constant through the 
Pagels-Stokar formula~\cite{Pagels-Stokar} at $T=\mu=0$.
Most results are presented by the ratio to the pion decay constant at
$T=\mu=0$.  When we present actual values, we use the value of the
pion decay constant in the chiral limit,
$f_\pi=88$\,MeV~\cite{Gasser-Leutwyler:SU(2)}. 
(This value of $f_\pi$ yields $\Lambda_{\rm qcd}\simeq582$\,MeV.)

\subsection{Preliminary}

Here we will summarize the framework of the numerical analysis done
below.  Variables $x$ and $y$ in Eq.~(\ref{xy}) take continuous values
from $0$ to $\infty$.  To solve the SDE numerically we first introduce
new variables $X$ and $Y$ as $X\equiv\ln(x/\Lambda_{\rm qcd})$ and
$Y\equiv\ln(y/\Lambda_{\rm qcd})$.  These variables take values from
$-\infty$ to $\infty$. Dominant contributions to the decay
constant and the 
effective
potential lie around $0$, {\it i.e.},
$x$ or $y$ is around $\Lambda_{\rm qcd}$, as shown later.
Then we introduce ultraviolet and infrared cutoffs, 
$X,Y\in[\lambda_{\rm IR},\lambda_{\rm UV}]$.
Finally, we descretize $X$ and $Y$ at $N_X$ points evenly:
\begin{eqnarray}
X_i, Y_i &=& 
\lambda_{\rm IR} + \Delta X \cdot i \ , 
\qquad i = 0 , 1 , \ldots , N_X - 1 \ ,
\nonumber\\
&&
\Delta X = \frac{\lambda_{\rm UV}-\lambda_{\rm IR}}{N_X-1} \ .
\label{def:XY}
\end{eqnarray}
Accordingly integrations over $x$ and $y$ are replaced with 
appropriate summations
\begin{equation}
\int dx \,,\ \int dy \ \rightarrow \ 
\Delta X \sum_i e^{X_i} \,,\ \Delta X \sum_i e^{Y_i} \ .
\label{int}
\end{equation}
We extract the derivative of $B(p)$ with respect to $x$, which is
used in the formula for $f_{\pi T}$ in Eq.~(\ref{fpi T}), from three
points by using a numerical differentiation formula:
\begin{equation}
x \left. \frac{\partial B(x)}{\partial x} 
\right\vert_{x/\Lambda_{\rm qcd}=e^{X_i}}
\simeq
\frac{4 B(x_{i+1}) - 3 B(x_i) - B(x_{i+2})}{2 \Delta X}
\ .
\end{equation}

In the case of $T=0$ Matsubara frequency sum becomes an integration
over $u$ or $v$ from $-\infty$ to $\infty$.  By using the property
(\ref{rel: B}) we can restrict this region from $0$ to $\infty$.  Then
we will perform a procedure similar to that for $x$ and $y$:
\begin{eqnarray}
&& 
U_n = \ln (u_n/\Lambda_{\rm qcd}) \,, \ 
V_n = \ln (v_n/\Lambda_{\rm qcd}) = 
\Lambda_{\rm IR} + \Delta U \cdot n \ ,
\quad n = 0 , 1, \ldots, N_U-1 \ ,
\nonumber\\
&& \qquad
\Delta U = \frac{\Lambda_{\rm UV} - \Lambda_{\rm IR}}{N_U-1} \ .
\label{def:UV}
\end{eqnarray}
In the case of $T\neq0$ we truncate the infinite sum of Matsubara
frequency at a finite number $N_U$:
\begin{equation}
\sum_{n=-\infty}^{\infty} \rightarrow 
\sum_{n=-N_U-1}^{N_U} \ .
\label{cut sum}
\end{equation}
The above  Eqs.~(\ref{def:XY})--(\ref{cut sum}) set regularizations
adopted in this paper.  We will check that the results are independent
of the regularizations.

Solving the SDE is done by an iteration method:
\begin{equation}
B_{\rm new}(p) = K(p,k) \ast 
\frac{B_{\rm old}(k)}{B^2_{\rm old}(k)-k^2} \ .
\end{equation}
Starting from a trial function, we stop the iteration if the
following condition is satisfied:
\begin{eqnarray}
\varepsilon \Lambda_{\rm qcd}^6 &>&
\frac{1}{4} \mbox{tr} 
\left[
  \left( \frac{\delta V}{\delta [S_F(p)]_{\rm old}} \right)^{\dag}
  \ast
  \left( \frac{\delta V}{\delta [S_F(p)]_{\rm old}} \right)
\right]
\nonumber\\
&=&
\left( B_{\rm old}(p) - B_{\rm new}(p) \right)^{\dag} \ast
\left( B_{\rm old}(p) - B_{\rm new}(p) \right) \ ,
\end{eqnarray}
with suitably small $\varepsilon$.
To obtain the second line we have used the fact that $A(p)-1=C(p)=0$
is an approximate solution.  This condition is natural in the sense
that the stationary condition of the effective potential is satisfied
within a required error.  In this paper we use
$\varepsilon=10^{-10}$.

\subsection{$T=0$ and $\mu\neq0$}

\begin{figure}[htbp]
\begin{center}
\epsfysize=3.0in
\ \epsfbox{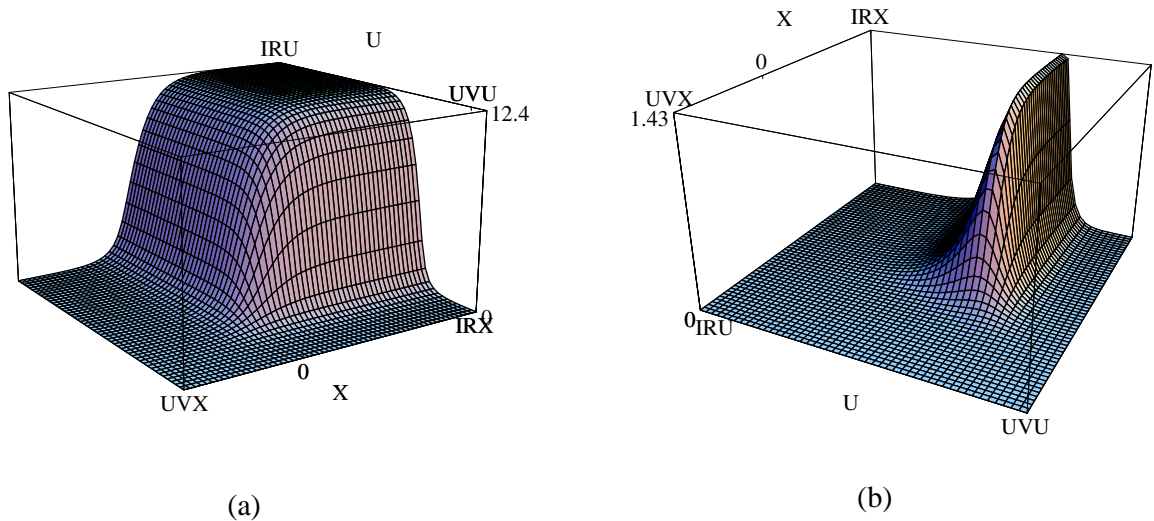}
\end{center}
\caption[]{Solution of the SDE at $\mu/f_\pi=0.3$ for $N_U=70$,
$N_X=60$: (a) real part; (b) imaginary part.  Scale 
$\Lambda_{\rm qcd}$ is indicated by $0$ on  
$U$ and $X$ axes.  Each number on the top of the vertical axis is the
maximum value of $\mbox{Re}B/f_\pi$ or $\mbox{Im}B/f_\pi$.
}\label{fig:sols}
\end{figure}

\begin{figure}[htbp]
\begin{center}
\epsfysize=3.0in
\ \epsfbox{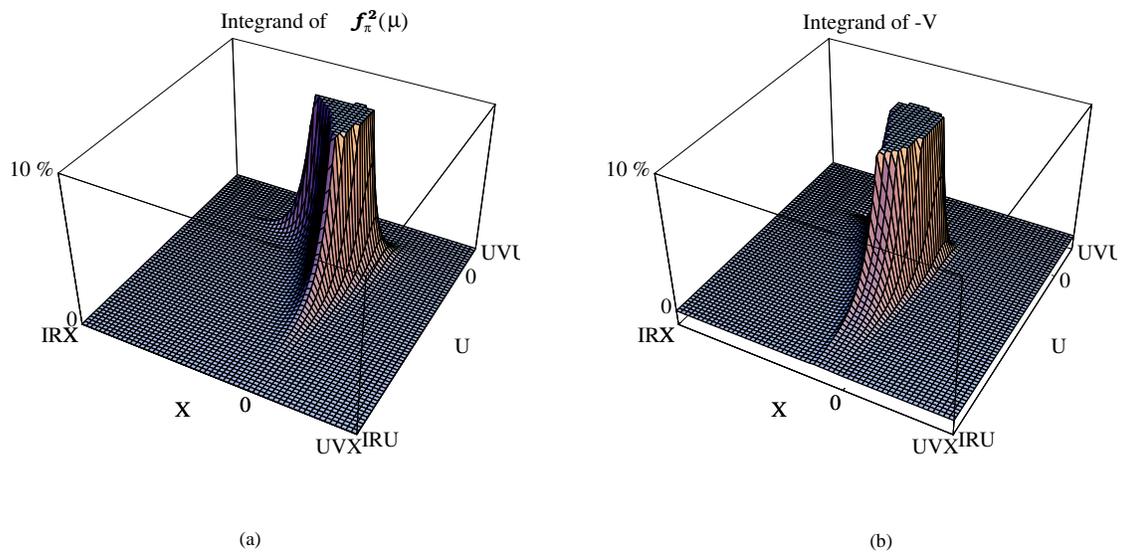}
\end{center}
\caption[]{Integrands of (a) $f_{\pi}^2(\mu)$ 
and (b) $-\overline{V}[B_{\rm sol}]$ at
$\mu/f_\pi=0.3$ for $N_U=70$, $N_X=60$.  The upper 9/10 of each figure
is clipped. 
}\label{fig:integrand}
\end{figure}

As we discussed in the previous subsection 
the
infinite Matsubara
frequency sum becomes an integration over continuous variable $u$
or $v$ at $T=0$.
Descretizations of these variables as well as $x$ and $y$ are done
as in Eqs.~(\ref{def:XY}) and (\ref{def:UV}).  We use
the following choice of infrared and ultraviolet cutoffs:
\begin{eqnarray}
X,\,Y &\in& \left[ \, -4.9 \, , \, 2.9 \, \right] \ , \nonumber\\
U,\,V &\in& \left[ \, -10.0 \, , \, 2.8 \, \right] \ .
\label{XYUV:range}
\end{eqnarray}
For initial trial functions we use the following two types:
\begin{eqnarray}
&\mbox{(A)}& \qquad B(p) = \left. B_{\rm sol}(p)\right\vert_{T=\mu=0} \ ,
\nonumber\\
&\mbox{(B)}& \qquad B(p) = 
\left\{\begin{array}{ll}
 0.1\cdot\Lambda_{\rm qcd} \ , 
   & \mbox{for}\ U_n \le U_{20} \ \mbox{and} \ X_i \le X_{10} \ , \\
 0   \ , & \mbox{otherwise} \ ,
\end{array}
\right.
\label{initial}
\end{eqnarray}
where $\left.B_{\rm sol}(p)\right\vert_{T=\mu=0}$ is a solution of the
SDE at $T=\mu=0$, and $U_n$ and $X_i$ are descretized variables as in
Eqs.~(\ref{def:XY}) and (\ref{def:UV}).  
To check validity of the cutoffs in Eq.~(\ref{XYUV:range})
we show (a) real part and (b) imaginary part of the solution in
Fig.~\ref{fig:sols}, and integrands of 
(a) $f_{\pi}^2(\mu)$, 
(b) $-\overline{V}[B_{\rm sol}]$ in Fig.~\ref{fig:integrand},
at $\mu/f_\pi = 3.0$ for $(N_U,N_X)=(70,60)$ by using the initial
trial function (A).
Figure~\ref{fig:sols}(a) 
shows that the real part becomes small above
$\Lambda_{\rm qcd}$.  This behavior is similar to that of the solution
at $T=\mu=0$.  The dependence on $x$ of the imaginary part is similar
to that of the real part.  Since the imaginary part is an odd function
in $u$, it becomes zero in the infrared region of $u$.
Figure~\ref{fig:integrand} shows that the
dominant part of each integrand lies within the integration range.
These imply that the choice of range in Eq.~(\ref{XYUV:range}) is
enough.

\begin{figure}[htbp]
\begin{center}
\epsfysize=3.0in
\ \epsfbox{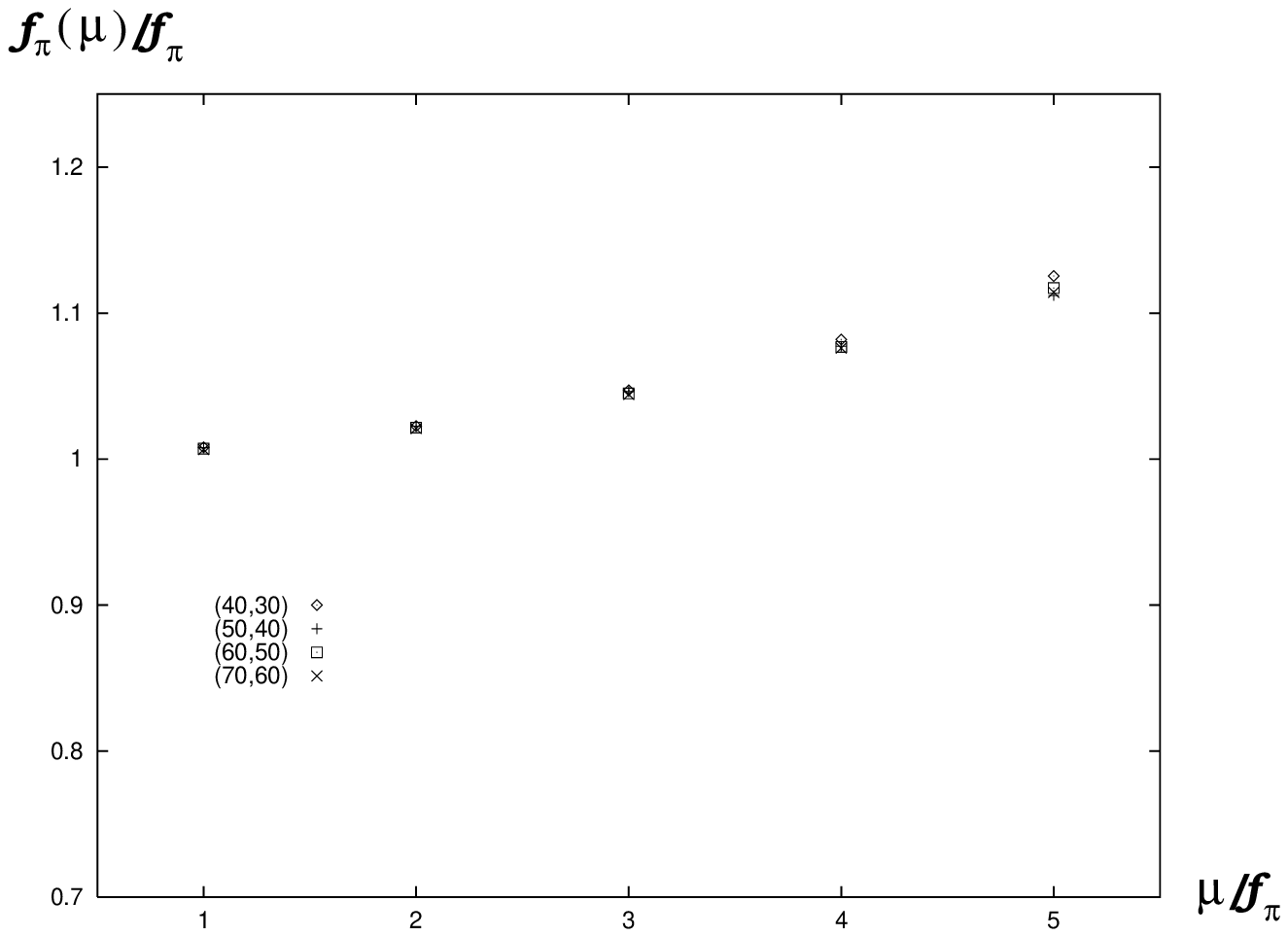}\\
\end{center}
\caption[]{Typical values of
$f_{\pi}(\mu)/f_\pi$ for four choices of the size of
descretization, $(N_U,N_X)=(40,30)$, $(50,40)$, $(60,50)$ and 
$(70,60)$.
}\label{fig:size dependence:f}
\end{figure}

\begin{figure}[htbp]
\begin{center}
\epsfysize=3.0in
\ \epsfbox{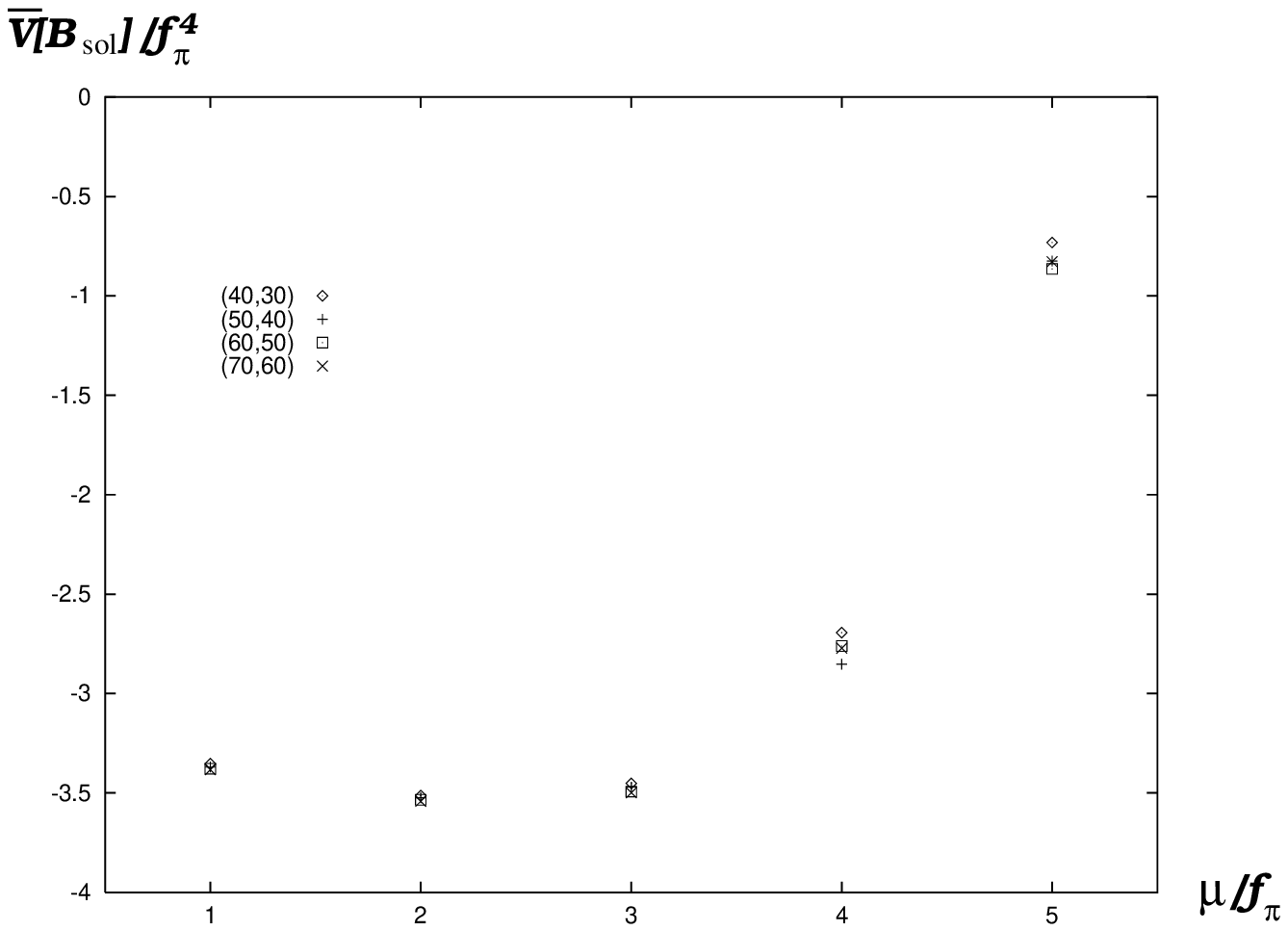}\\
\end{center}
\caption[]{Typical values of
$\overline{V}[B_{\rm sol}]/f_\pi^4$ for four choices of the size of
descretization, $(N_U,N_X)=(40,30)$, $(50,40)$, $(60,50)$ and 
$(70,60)$.
}\label{fig:size dependence:v}
\end{figure}

Next let us study the dependence of the results on the size of
descretization.  We show typical values of $f_{\pi}(\mu)/f_\pi$
in Fig.~\ref{fig:size dependence:f} 
and that of $\overline{V}[B_{\rm sol}]/f_\pi^4$ in 
Fig.~\ref{fig:size dependence:v} for four choices of the size of
descretization, $(N_U,N_X)=(40,30)$, $(50,40)$, $(60,50)$ and
$(70,60)$.  This shows that the choice $(N_U,N_X)=(70,60)$ is large
enough for the present purpose.

\begin{figure}
\begin{center}
\epsfysize=3.0in
\ \epsfbox{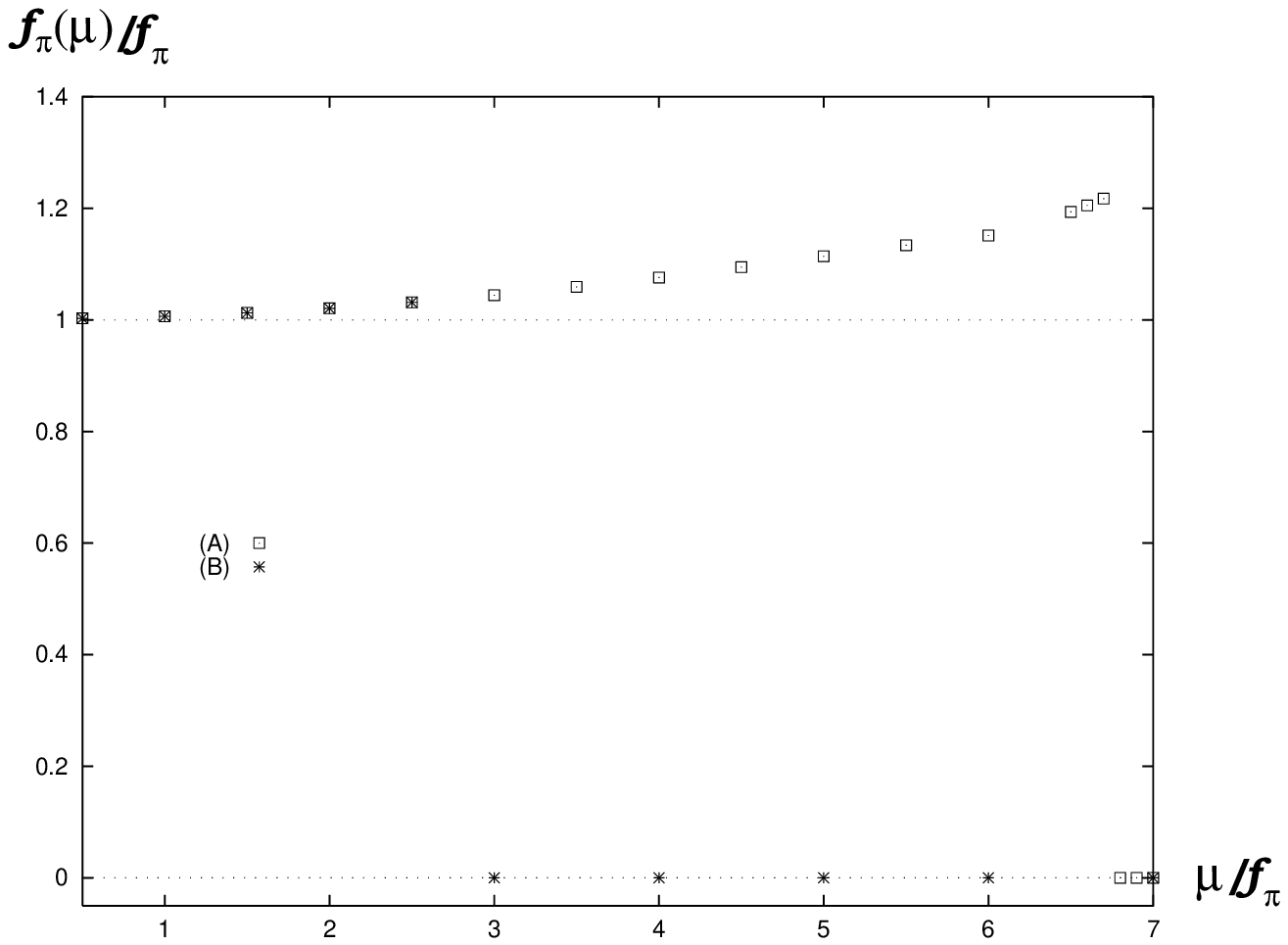} \\
\end{center}
\caption[]{
$\mu$-dependence of $f_{\pi}(\mu)/f_\pi$ at $T=0$ for 
two choices of the initial trial functions (A) and (B)
in Eq.~(\ref{initial}).
}\label{fig:5}
\end{figure}

Now we show the resultant values of $f_{\pi}(\mu)/f_\pi$
for two choices of initial trial functions in 
Fig.~\ref{fig:5}.
Below $\mu/f_\pi=3.0$ both the trial functions converge to the same
non-trivial solution.  The value of $f_{\pi}(\mu)$ increases slightly.
The choice (B) in Eq.~(\ref{initial}) converges to the trivial
solution above $\mu/f_\pi=3.0$.
However, the trial function (A) converges to a non-trivial solution,
and the resultant value of $f_\pi(\mu)$ increases.  Above
$\mu/f_\pi=6.8$ (A) also converges to the trivial 
solution.
In the range $3.0\le\mu/f_\pi\le6.8$ two initial trial functions
converge to different solutions: one corresponds to the chiral broken
vacuum and another to the symmetric vacuum.  Moreover, the trivial
solution is always a solution of the SDE (\ref{SD: B}).  Then we have
to study which of the vacua is the true vacuum.

\begin{figure}[htbp]
\begin{center}
\epsfysize=3.0in
\ \epsfbox{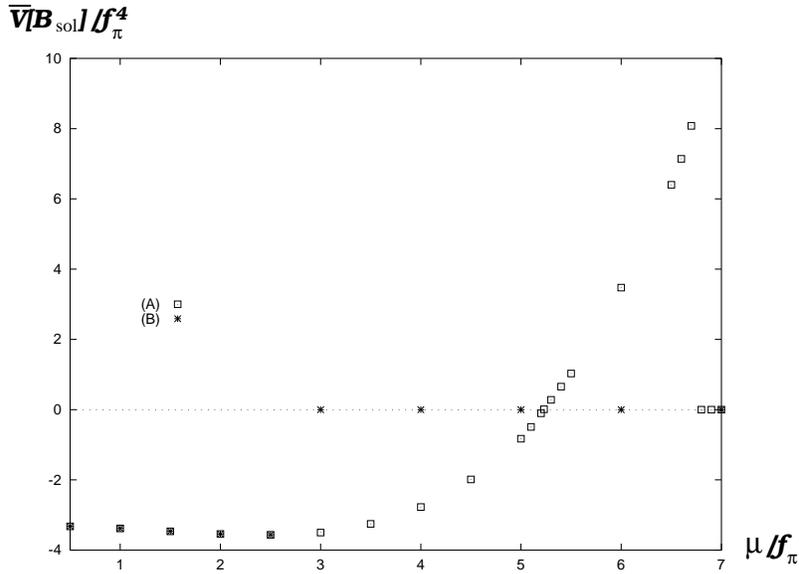}\\
\end{center}
\caption[]{
$\mu$-dependence of the effective potential 
$\overline{V}[B_{\rm sol}]/f_\pi^4$ at $T=0$.
}\label{fig:6}
\end{figure}

As we discussed in Sect.~\ref{sec:2}, the true vacuum is determined by
studying the effective potential.
We show the value of the effective potential 
$\overline{V}[B_{\rm sol}]/f_\pi^4$ in Fig.~\ref{fig:6}.
Since the value for the trivial solution is already subtracted from
the expression in Eq.~(\ref{eff pot sol}), positive (negative) value
of $\overline{V}[B_{\rm sol}]$ implies that the energy of the chiral
broken vacuum is larger (smaller) than that of the symmetric vacuum.
For $3.0\le\mu/f_\pi<5.23$, 
$\overline{V}[B_{\rm sol}]_{(A)} < \overline{V}[B_{\rm sol}]_{(B)}
=0$, which implies that the chiral broken vacuum is energetically
favored.  On the other hand, for $5.23\le\mu/f_\pi\le6.8$, 
$\overline{V}[B_{\rm sol}]_{(A)} > 0$, and the chiral symmetric vacuum
is the true vacuum.
The chiral symmetry is restored at the point where the value of
$\overline{V}[B_{\rm sol}]$ becomes positive: the chiral
phase transition occurs at $\mu=460$\,MeV ($\mu/f_\pi=5.23$).  
Since the value of the pion decay constant vanishes discontinuously at
that point, the phase transition is clearly of first order.  The
non-trivial solutions for $5.23\le\mu/f_\pi\le6.8$ correspond to
meta-stable states, which were shown in Ref.~\cite{BCDGP:PRD41} by
assuming a momentum dependence of the mass function.
The result here obtained by solving the SDE (\ref{SD: B}) agrees
qualitatively with their result.

\subsection{$T\neq0$ and $\mu=0$}

\begin{figure}[tbtp]
\begin{center}
\epsfysize=3.0in
\ \epsfbox{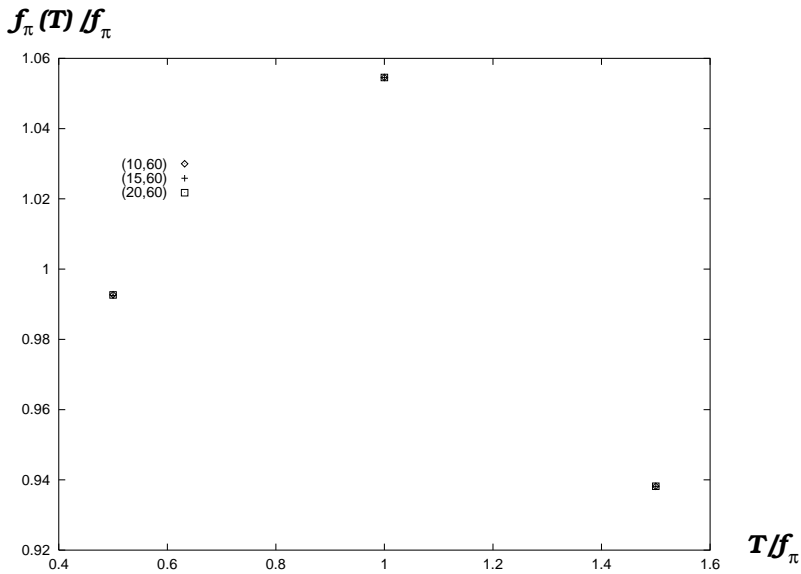}
\end{center}
\caption[]{
Values of $f_{\pi }(T)/f_\pi$ at
$T/f_\pi=0.5$, $1.0$ and $1.5$ for three choices of the truncation
point: $N_U=10$, $15$ and $20$.
}\label{fig:7}
\end{figure}

At non-zero temperature we perform Matsubara frequency sum by
truncating it at some finite number.  Integration over the spatial
momentum is done as shown in Eqs.~(\ref{def:XY}) and (\ref{int}). 
We have checked that the integration range given in the first line of
Eq.~(\ref{XYUV:range}) is large enough for the present purpose.  
Since the trial function (B) converges to the same solution as (A) for
small $\mu$ in the case of $T=0$, it may be enough to use the trial
function (A) for studying phase structure.  Here and henceforth we use
the trial function of type (A) only.
First, we check dependences of the results on the truncation.  In
Fig.~\ref{fig:7} we show the values of $f_{\pi}(T)/f_\pi$ at
$T/f_\pi=0.5$, $1.0$ and $1.5$ for three choices of the truncation
point: $N_U=10$, $15$ and $20$.
It is clear that the choice $N_U=20$ is large enough as the truncation
point for the present purpose.

\begin{figure}[tbtp]
\begin{center}
\epsfysize=3.1in
\ \epsfbox{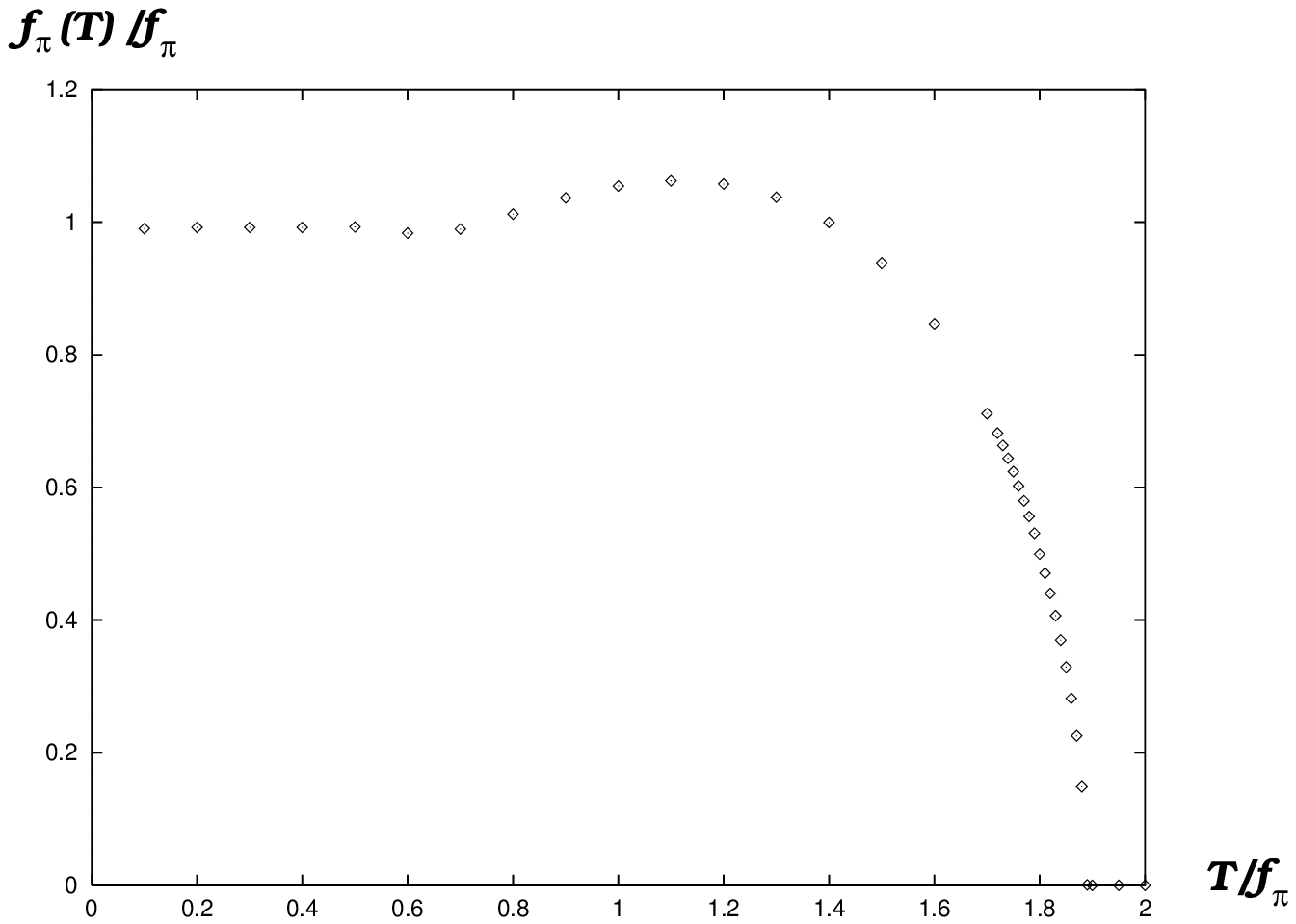}\\
\ (a)\\
\ \\
\epsfysize=3.1in
\ \epsfbox{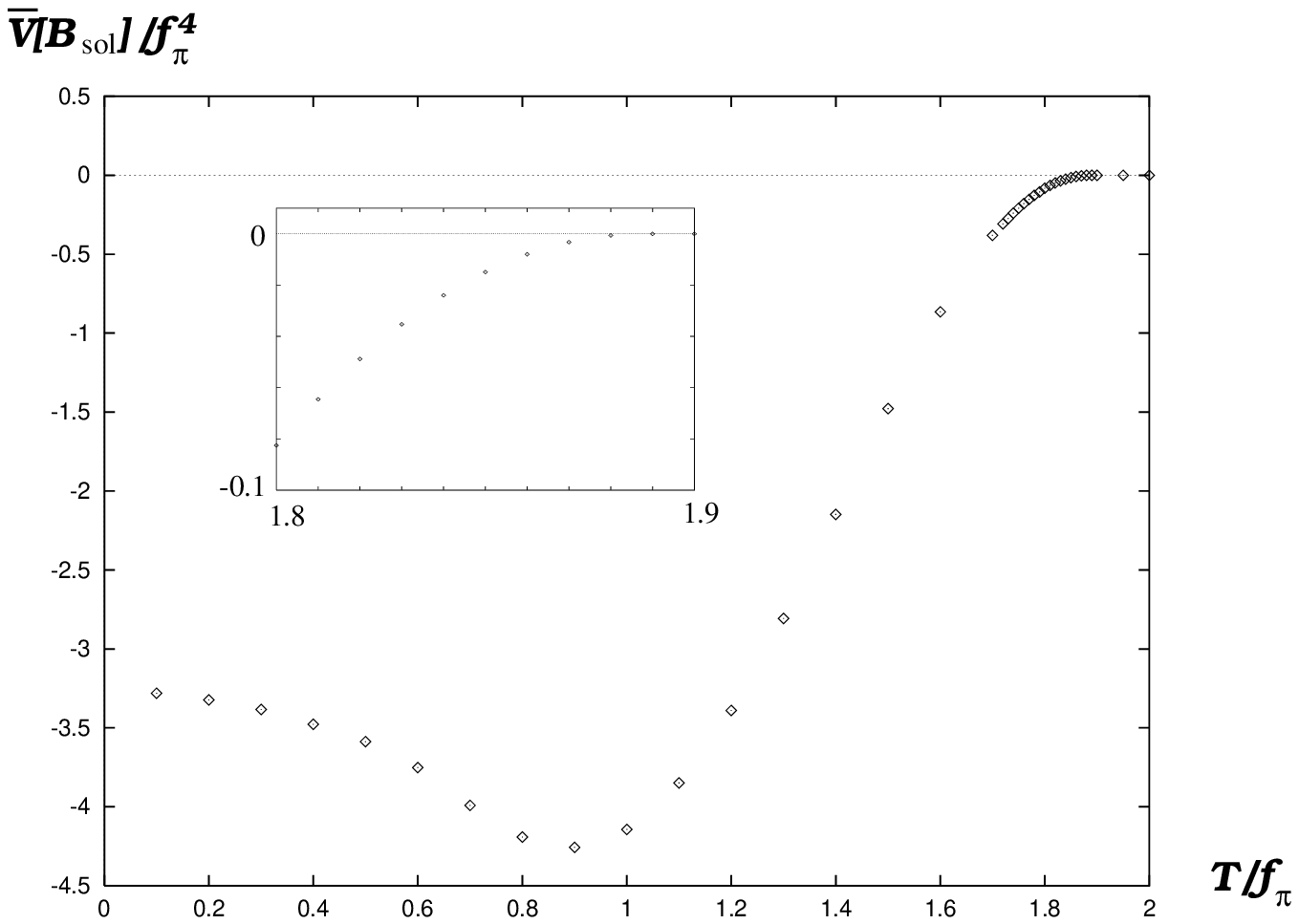}\\
\ (b)\\
\end{center}
\caption[]{
Temperature dependences of (a) $f_{\pi}(T)/f_\pi$
and (b) $\overline{V}[B_{\rm sol}]/f_\pi^4$.
}\label{fig:8}
\end{figure}

Resultant temperature dependences of $f_{\pi }(T)/f_\pi$ and
$\overline{V}[B_{\rm sol}]/f_\pi^4$ are shown in Fig.~\ref{fig:8}.
The value of $f_{\pi}(T)$ does not change in low temperature region.
It once increases around $T/f_\pi\sim0.7$ and decreases to zero above
that point, and finally reach to zero around $T/f_\pi=1.89$.
Differently with the previous case the value of the potential reaches
to zero around the temperature where the decay constant vanishes.
Since there are numerical errors in the present analysis, we can not
clearly show whether the decay constant and the potential vanish
simultaneously.  Our result shows that the chiral phase transition is
of second order or of very weak first order, and that the critical
temperature is around $T_c = 166$\,MeV.

\subsection{$T\neq0$ and $\mu\neq0$}

Now, we solve the SDE when both $T$ and $\mu$ are non-zero.  Infrared
and ultraviolet cutoffs for $x$-integration are fixed as in the first
equation of Eq.~(\ref{XYUV:range}), and the sizes of descretizations
are fixed to be $(N_U,N_X)=(20,60)$.

\begin{figure}[hbtp]
\begin{center}
\epsfysize=3.0in
\ \epsfbox{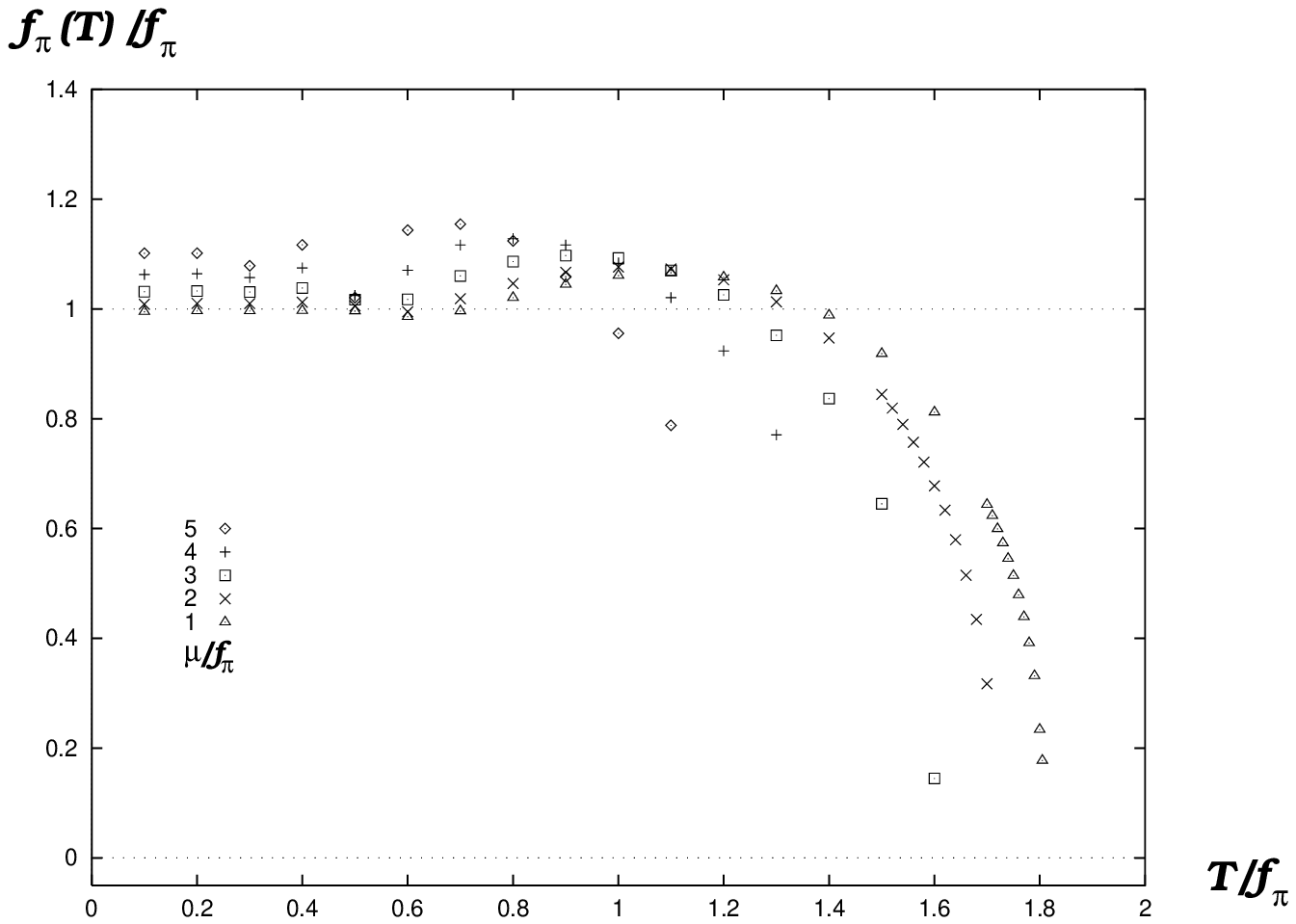}
\end{center}
\caption[]{
Temperature dependences of 
$f_{\pi L}/f_\pi$ for $\mu/f_\pi=1$, $2$, $3$, $4$ and $5$.
}\label{fig:fLT1}
\end{figure}

\begin{figure}[hbtp]
\begin{center}
\epsfysize=3.0in
\ \epsfbox{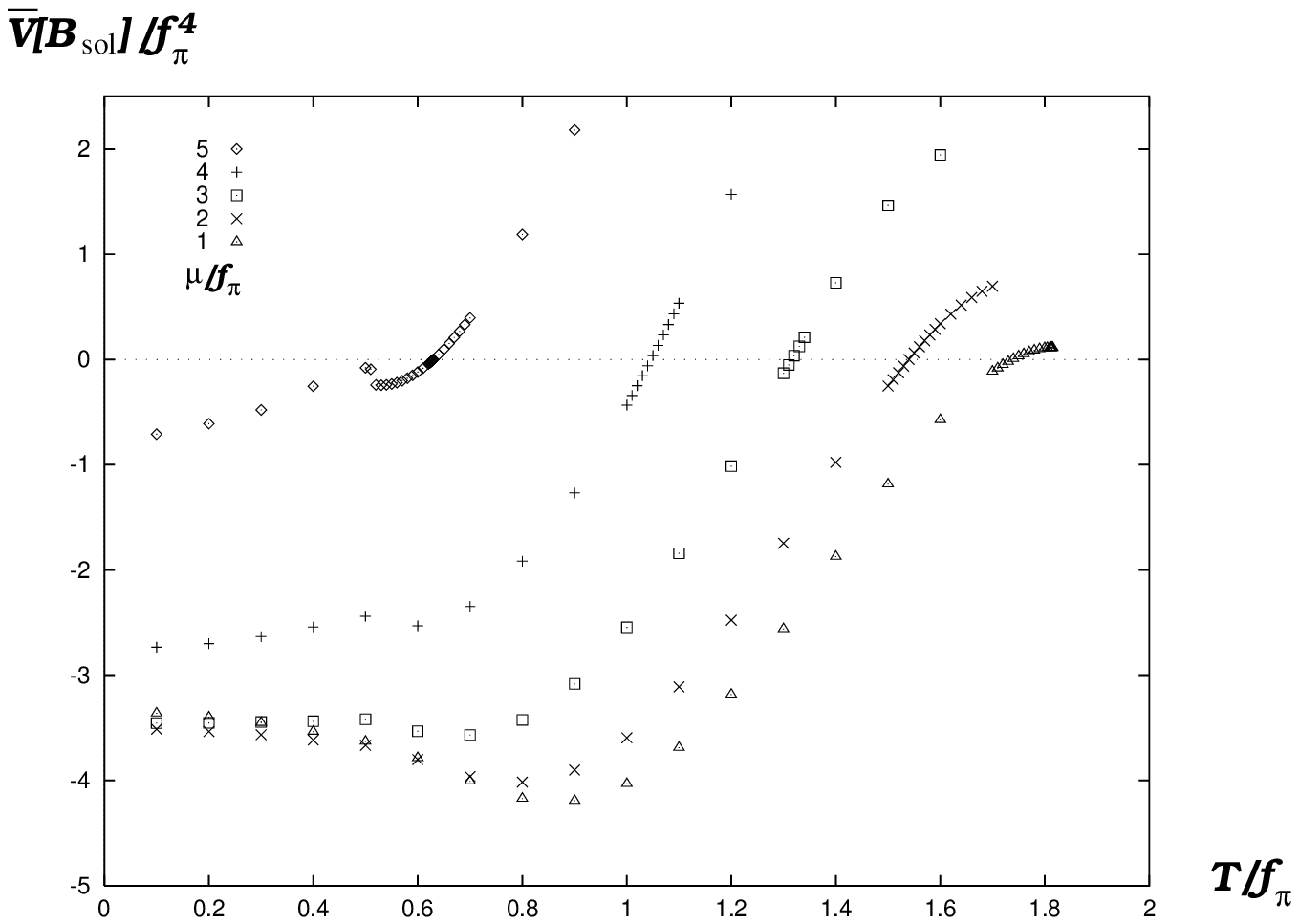}
\end{center}
\caption[]{
Temperature dependence of the effective potential 
$\overline{V}[B_{\rm sol}]/f_\pi^4$ for $\mu/f_\pi=1$, $2$, $3$, $4$
and $5$.
}\label{fig:v1}
\end{figure}

\begin{figure}[hbtp]
\begin{center}
\epsfysize=3.1in
\ \epsfbox{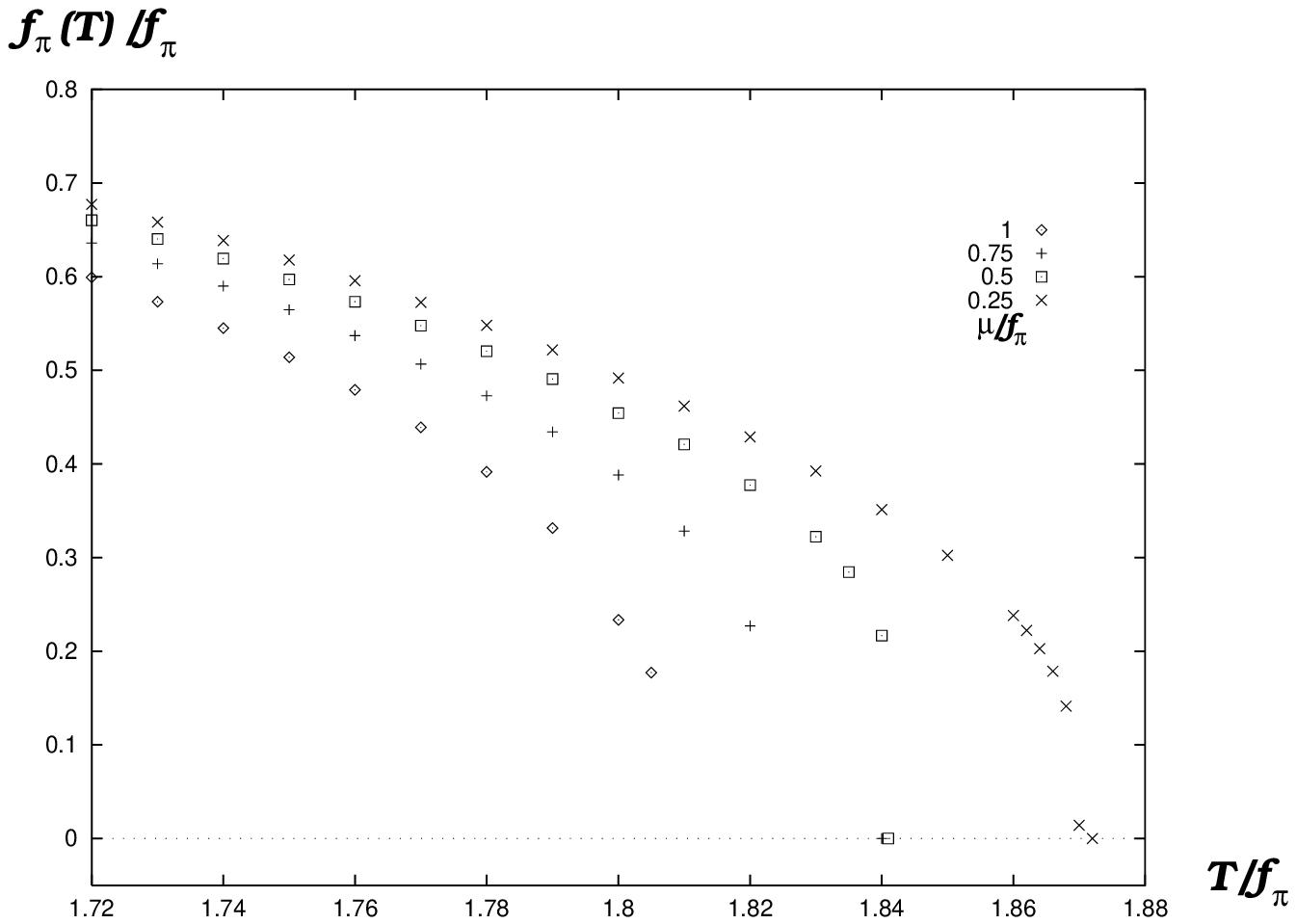}
\end{center}
\caption[]{
Temperature dependences of $f_{\pi}(T)/f_\pi$ 
for $\mu/f_\pi=0.25$, $0.5$, $0.75$ and $1.0$.
}\label{fig:fLT2}
\end{figure}

\begin{figure}[hbtp]
\begin{center}
\epsfysize=3.0in
\ \epsfbox{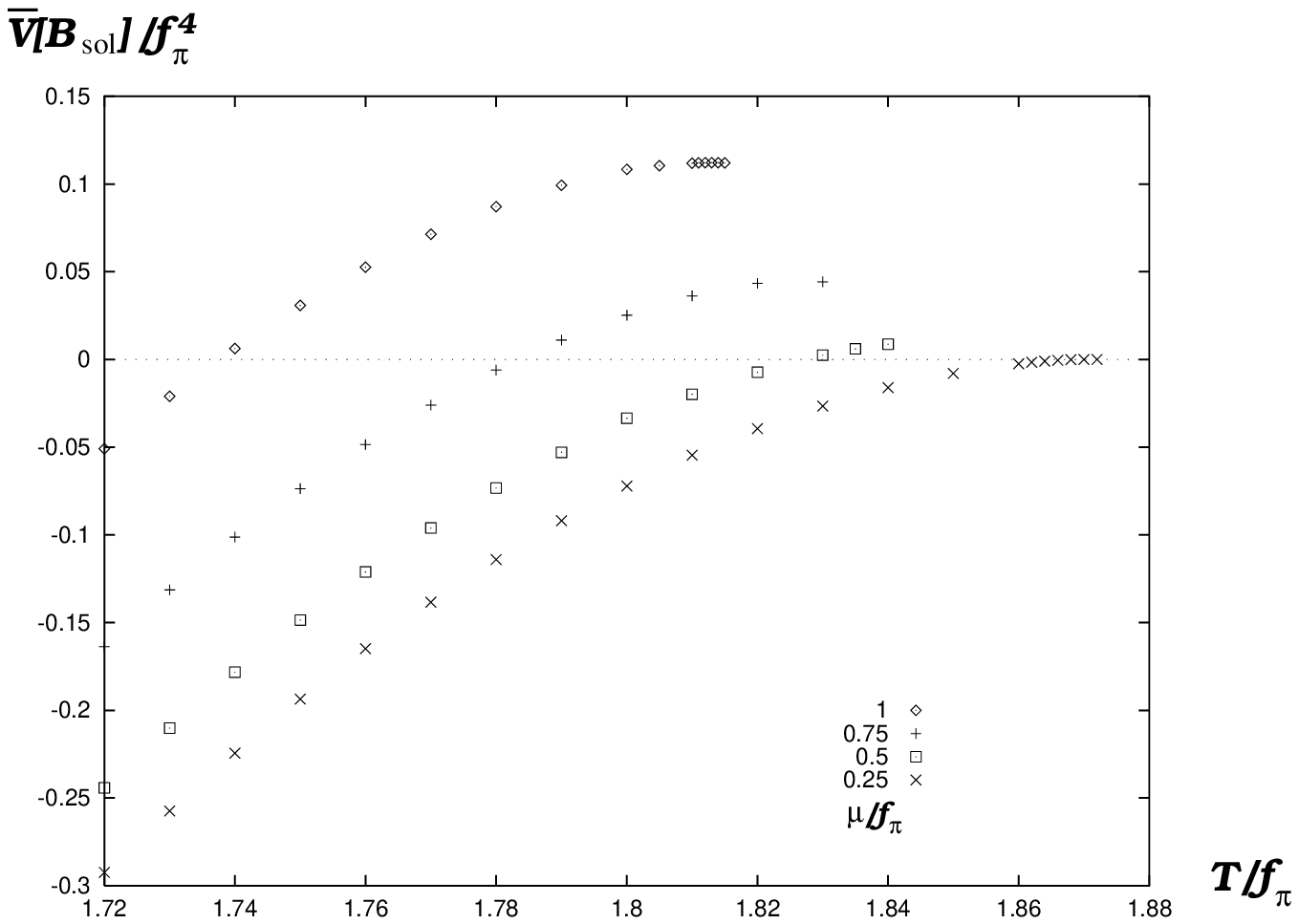}
\end{center}
\caption[]{
Temperature dependence of the effective potential 
$\overline{V}[B_{\rm sol}]/f_\pi^4$
for $\mu/f_\pi=0.25$, $0.5$, $0.75$ and $1.0$.
}\label{fig:v2}
\end{figure}

In the previous subsections we found that the phase transition at
$T=0$ and $\mu\neq0$ is of first order, while that at $T\neq0$ and
$\mu=0$ is of second order or of very weak first order.
Then one can expect phase transitions of first order for small
$T$ and large $\mu$, and those of weak first order for large $T$ and
small $\mu$.  First, we show the temperature dependence of 
$f_{\pi}(T)/f_\pi$ in Fig.~\ref{fig:fLT1} and that of 
$\overline{V}[B_{\rm sol}]/f_\pi^4$ in Fig.~\ref{fig:v1}
for $\mu/f_\pi=1$, $2$, $3$, $4$ and $5$.
Figure~\ref{fig:fLT1} shows that in all the cases the values of 
the pion decay constant once increase around $T\sim f_\pi$ and
decrease above that.
These, especially for $\mu/f_\pi = 1$ and $2$, behave as if the phase
transitions are of second order.
However, it is clear from Fig.~\ref{fig:v1}
that the values of the effective potential 
$\overline{V}[B_{\rm sol}]$ become positive before the values of 
the pion decay constant vanish. 
Then the phase transitions are clearly of first order as in the case
of $T=0$ and $\mu\neq0$. 
To study phase transitions for smaller $\mu$ we concentrate on 
temperature dependences around phase transition points.
Shown in Figs.~\ref{fig:fLT2} and \ref{fig:v2} are temperature
dependences of the pion decay constant and the effective potential for
$\mu/f_\pi=0.25$, $0.5$ and $0.75$ together with those for
$\mu/f_\pi=1$.
For $\mu/f_\pi=0.5$, $0.75$ and $1.0$ the values of the pion
decay constant approach to zero as if the phase transitions are
of second order.  However, the values of the effective
potential become positive before the values of the decay constant
reaches to zero.
Then we conclude that the phase transitions for $\mu/f_\pi=0.5$,
$0.75$ and $1.0$ are clearly of first order.  On the other hand, the
phase transition for $\mu/f_\pi=0.25$ is of second order or of very
weak first order.

\begin{figure}[hbtp]
\begin{center}
\epsfysize=3.0in
\ \epsfbox{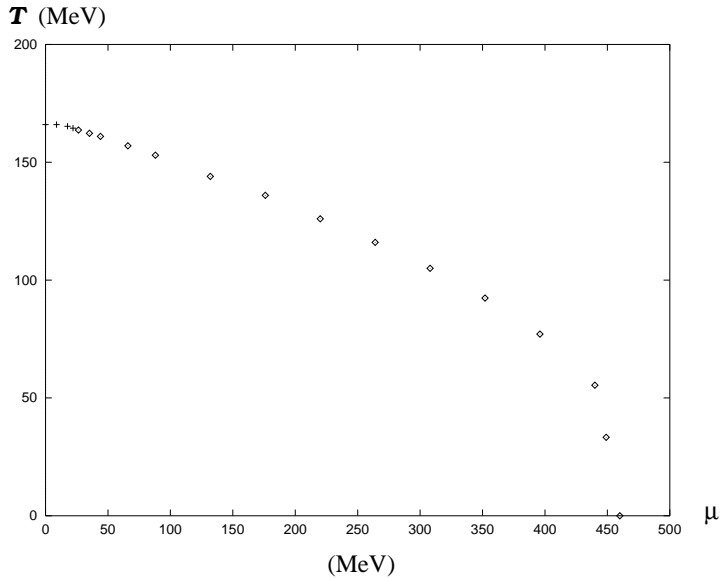}
\end{center}
\caption[]{
Phase diagram obtained by the present analysis.
Points indicated by $\diamond$ are the phase transition points of
first order, and points by $+$ of second order or very weak first
order.
}\label{fig:phase}
\end{figure}

Finally we show a phase diagram derived by the present analysis
in Fig.~\ref{fig:phase}.
As was expected, phase transitions for small $T$ and large $\mu$
are of first order, and those for large $T$ and small $\mu$ are of
second order or of very weak first order.

\section{Summary and Discussion}
\label{sec:discussion}

We analyzed the phase structure of QCD at finite temperature and
density by solving the self-consistent Schwinger-Dyson equation for
the quark propagator with the improved ladder approximation in the 
Landau gauge.  A pion decay constant was calculated by using a
generalized version of the Pagels-Stokar formula.  Chiral phase
transition point was determined by an effective potential for the
propagator.  When we raised the chemical potential $\mu$ with $T=0$,
the value of the effective potential for the broken vacuum became
bigger than that for the symmetric vacuum before the value of the pion
decay constant vanished.  The phase transition is clearly of first
order.  On the other hand, for $T\neq0$ and $\mu=0$ the value of the
effective potential for the broken vacuum reached to that for the
symmetric vacuum around the temperature where the value of the pion
decay constant vanished.  The phase transition is of second order or
very weak first order.  We presented the resultant phase diagram on
the general $T-\mu$ plane.  Phase transitions are clearly of first
order in most cases, and for small $\mu$ they are of second order or
very weak first order.  Our results show that it is important to use
the effective potential to study the phase structure at finite
temperature and density.

Finally, some comments are in order.
Generally the running coupling should include the term from 
vacuum polarization of quarks and gluons at finite temperature and
density. 
Moreover, the gluons at high temperature acquire an electric
screening mass of order $gT$.~\cite{KK85}
We dropped these effects, and used a running coupling and a gluon
propagator of the same forms at $T=\mu=0$.
(The same approximation was used in Ref.~\cite{Taniguchi-Yoshida}.)
These effects can be included by using different running coupling and
gluon propagator which depend explicitly on $T$ and $\mu$ such as 
in Refs.~\cite{BCDGP:PRD41,KY95}.

The approximation of $A-1=C=0$ might not be good for high tempreture
and/or density.  However, inclusion of the deviations of $A-1$ and $C$
from zero requires the large number for truncating the Matsubara
frequency, and it is not efficient to apply the method used in this
paper.
New method to solve the SDE may be needed.
We expect that the inclusion of the deviations of $A-1$ and $C$ from
zero does not change the structure of the phase transition
shown in the present paper.

\section*{Acknowledgement}

We would like to thank Prof.~Paul Frampton,
Prof.~Taichiro Kugo, Prof.~Jack Ng, Prof.~Ryan Rohm and 
Prof.~Joe Schechter for useful discussions and comments.

\end{document}